\documentclass[11pt ]{article}
\usepackage{amsmath}
\usepackage{graphicx}
\usepackage{hyperref}
\usepackage{lmodern}
\usepackage{amssymb}
\usepackage{booktabs}
\usepackage{tikz}
\usetikzlibrary{patterns}

\makeatletter
\def\eqnarray{ \stepcounter{equation} \let\@currentlabel=\theequation
 \global\@eqnswtrue
 \global\@eqcnt\z@
 \tabskip\@centering
 \let\\=\@eqncr
 $$\halign to \displaywidth\bgroup\@eqnsel\hskip\@centering
 $\displaystyle\tabskip\z@{##}$&\global\@eqcnt\@ne
 \hfil$\displaystyle{{}##{}}$\hfil
 &\global\@eqcnt\tw@$\displaystyle\tabskip\z@{##}$\hfil
 \tabskip\@centering&\llap{##}\tabskip\z@\cr}
\makeatother

\makeatletter
\def\@arrayacol{\edef\@preamble{\@preamble \hskip .5\arraycolsep}}
\def\array{\let\@acol\@arrayacol \let\@classz\@arrayclassz
\let\@classiv\@arrayclassiv \let\\\@arraycr\def\@halignto{}\@tabarray}
\makeatother

\makeatletter
\newcounter{subeqncnt}
\def\thesubeqncnt{\alph{subeqncnt}}
\def\subequations{\begingroup%
   \stepcounter{equation}\edef\@tempa{\theequation}%
   \let\c@equation\c@subeqncnt\c@subeqncnt\z@
   \edef\theequation{\@tempa\noexpand\thesubeqncnt}}

\makeatother

\newcommand{\be}{\begin{equation}}
\newcommand{\ee}{\end{equation}}

\newcommand{\beqa}{\begin{eqnarray}}
\newcommand{\eeqa}{\end{eqnarray}}
\newcommand{\nn}{\nonumber}

\def\CD {{\cal D}}

\def\CF {{\cal F}}
\def\CG {{\cal G}}

\def\CL {{\cal L}}

\def\Det{{\rm Det}}

\begin{document}

\setlength{\baselineskip}{7mm}
\begin{titlepage}
\begin{flushright}
{\tt NRCPS-HE-22-2023} \\
March, 2023
\end{flushright}

\vspace{1cm}

\begin{center}
{\it \Large 
Yang-Mills Effective Lagrangian \\
\vspace{1cm}
Contribution of  Leutwyler Zero Mode Chromons \\ 

}

\vspace{1cm}

{ {George~Savvidy  }}\footnote{savvidy(AT)inp.demokritos.gr}

\vspace{1cm}

 {\it Institute of Nuclear and Particle Physics, NCSR Demokritos, GR-15310 Athens, Greece}\\

\end{center}

\vspace{1cm}

\begin{abstract}

Recently we calculated the contribution of Leutwyler zero mode chromons to the Yang-Mills effective Lagrangian by integrating over collective coordinates of the nonlinearly interacting zero modes.  Here we analyse the behaviour of the partition function prefactor and the integration measure Jacobian  of the zero modes and show that these factors do not contribute to the effective Lagrangian.

\end{abstract}

\end{titlepage}

{\it  {\bf 1} Introduction.}
The (anti)self-dual  covariantly-constant vacuum fields\footnote{Here, and afterwards, the phrase  "vacuum fields" refers to the gauge fields that are the solutions of the sourseless Yang-Mills equation.}   have positive/stable and infinitely many zero modes, so called Leutwyller zero mode chromons \cite{ Leutwyler:1980ev, Leutwyler:1980ma, Minkowski:1981ma}.  Earlier we integrated over the collective coordinates of the nonlinearly interacting zero modes and obtained their contribution to the effective Lagrangian \cite{Savvidy:2022jcr}. Here we will analyse the behaviour of the corresponding partition function prefactor and the collective coordinate integration measure Jacobian of the zero mode chromons and will show that these factors do not contribute to the effective Lagrangian and do not alter the earlier results \cite{Savvidy:1977as, Matinyan:1976mp, Batalin:1976uv, PhDTheses,Savvidy:2019grj}.  Our method of the integration over zero mode coordinates is a generalisation of the t'Hooft integration over the instanton zero mode collective coordinates  \cite{tHooft:1976snw} and  is reminiscent  to the integration technique developed  by Feynman \cite{Feynman:1965} and  Fujikawa \cite{Fujikawa:1979ay}. \\
{\it {\bf 2} Euclidean Path Integral.} 
The  effective Lagrangian \cite{Heisenberg:1935qt,  Schwinger:1951nm,Coleman:1973jx}
  in the Yang-Mills theory \cite{Savvidy:1977as, Matinyan:1976mp, Batalin:1976uv, PhDTheses,Savvidy:2019grj} can be analysed in the  Euclidean space  \cite{tHooft:1976snw, Coleman} and has the following form \cite{Leutwyler:1980ev,  Leutwyler:1980ma,Savvidy:2019grj,Coleman,Flory:1983td,parthasarathy, Kay1983,kumar}:
 \beqa\label{euclidfield}
 \int^{ \vec{A}^{~'}}_{\vec{A}} \CD \vec{A}(\tau) e^{-S_E[A(\tau)]} \equiv N_f  e^{-\int  d^4 x \CL_{eff}(\vec{A}^{'}, \vec{A})},~~~
\eeqa
where  the path integral involves the integration over  the Euclidean field  configurations $A^{a}_{i}( \vec{x}, \tau )$ that have  the proper boundary values 
$
A^{a}_{i}(\vec{x},0) = A^{a}_{i}(\vec{x})$,~  $A^{a}_{i}(\vec{x},T) = A^{'a}_{i}(\vec{x}).
$
The Euclidean action is:
\be\label{action}
S_E= {1\over 4} \int d^4 x G^{a}_{\mu\nu}G^{a}_{\mu\nu},
\ee
where $x_0 \rightarrow -ix_0=-i \tau$,~$A_0 \rightarrow i A_0$ and $G^{a}_{\mu\nu}(A) = \partial_{\mu} A^{a}_{\nu} - \partial_{\nu} A^{a}_{\mu} +g f^{abc} A^{b}_{\mu}A^{c}_{\nu}
$. 
The real electric field corresponds to an imaginary electric field in Euclidean  space $ \vec{E} \rightarrow  i \vec{E}$.  The behaviour of vacuum fields  in the vicinity of a given interpolating-vacuum field $B^a_{\mu}(x)$ that starts at $A^{a}_{i}(\vec{x})$ and ends at $A^{'a}_{i}(\vec{x})$ is represent in the following form:
$
A^{a}_{\mu}(x) = B^a_{\mu}(x) + a^a_{\mu}(x),
$
where quantum field $a^a_{\mu}(x)$ has the zero boundary values.
The field strength  $G^a_{\mu\nu}(A)$  can be expanded in   $a^a_{\mu}(x)$  as 
$G^a_{\mu\nu}(A) = G^a_{\mu\nu}(B) + \nabla^{ab}_{\mu} a^{b}_{\nu}  - \nabla^{ab}_{\nu} a^{b}_{\mu}+ g f^{abc} a^{b}_{\mu}a^{c}_{\nu}, 
$
where $\nabla^{ab}_{\mu} a^{b}_{\nu} = \partial_{\mu} a^{a}_{\nu} +g f^{abc} B^{b}_{\mu}a^{c}_{\nu}$ is a covariant derivative with respect to the vacuum  field $B^a_{\mu}(x)$ and $[\nabla_{\mu},\nabla_{\nu}]^{ab} = g f^{acb} G^{c}_{\mu\nu}(B)$. For the Euclidean action (\ref{action})  we will have:
\beqa\label{fluctuationall}
S_E&=&{1\over 4}\int d^4x G^a_{\mu\nu}(B) G^a_{\mu\nu}(B) + \int  d^4x G^a_{\mu\nu}(B) \nabla^{ab}_{\mu} a^{a}_{\nu}+\nn\\
&+& {1\over 4}\int  d^4 x \Big((\nabla^{ab}_{\mu} a^{b}_{\nu} -\nabla^{ab}_{\nu} a^{b}_{\mu}+ g f^{abc} a^{b}_{\mu}a^{c}_{\nu})^2 + 2 g G^a_{\mu\nu}(B) f^{abc} a^{b}_{\mu}a^{c}_{\nu} \Big).
\eeqa
The vacuum field $B^a_{\mu}$ is supposed to satisfy the Euclidean Yang-Mills equation:
\be\label{YMeq}
\nabla_{\mu} G_{\mu\nu}(B)=0, 
\ee
so that the second term in (\ref{fluctuationall}) that is linear in $a^a_{\mu}(x)$ vanishes and the effective Lagrangian is a gauge invariant functional \cite{Batalin:1976uv}.   The quadratic part of the action has the following form:
\be\label{linear}
K_E= {1\over 4}\int d^4x \Big[(\nabla^{ab}_{\mu} a^{b}_{\nu} -\nabla^{ab}_{\nu} a^{b}_{\mu} )^2 + 2 g G^a_{\mu\nu}(B) f^{abc} a^{b}_{\mu}a^{c}_{\nu} \Big].
\ee
The nonlinear terms in the Euclidean Lagrangian (\ref{fluctuationall})  are cubic and quartic in $a^a_{\mu}(x)$:
\be\label{nonlinear}
V_E=\int d^4x \Big(- g f^{abc} a^{b}_{\nu}a^{c}_{\mu} \nabla^{ad}_{\mu} a^{d}_{\nu} + {g^2 \over 4} (f^{abc}a^{b}_{\mu}a^{c}_{\nu})^2\Big) .
\ee
The boundary fields are taken to be equal  $A^{a}_{i}(\vec{x})=A^{'a}_{i}(\vec{x})$, to be the {\it covariantly-constant} fields \cite{Duff:1975ue,  Batalin:1976uv, PhDTheses} and obey the 
equation (\ref{YMeq}).  Then the vacuum  field $B^a_{\mu}(x)$ will has the following form:
\be\label{interpol}
B^a_{\mu}(x) = -{1\over 2} F_{\mu\nu} x_{\nu} \delta^{a}_{3},   
\ee
where the vectors $E_{i }$ and $H_i$ can be alined with the z-axis by two independent rotations  \cite{Leutwyler:1980ev}:
\be\label{backfields}
F_{12}=H,~~~F_{30}= E.
\ee
The Euclidean gauge invariants are: $\CF_E= {1\over 4} G^a_{\mu\nu}G^a_{\mu\nu} ={H^2 +E^2\over 2}$ and the $\CG_E ={1\over 4} G^a_{\mu\nu} \tilde{G}^a_{\mu\nu} = H E$.    In the quadratic approximation (\ref{linear})  the stability of  vacuum fields  reduces to the following  eigenvalue problem \cite{Leutwyler:1980ev,  Leutwyler:1980ma, Savvidy:2022jcr}:
 \be\label{quadratic}
 -\nabla^{ab}_{\mu}( \nabla^{bc}_{\mu} a^{c}_{\nu} - \nabla^{bc}_{\nu} a^{c}_{\mu}) + g f^{abc} G^b_{\mu\nu} a^{c}_{\mu} = \lambda a^{a}_{\nu}.
 \ee
It is convenient to decompose the field $a_{\mu}$ into the neutral and charged components\footnote{We will consider the $SU(2)$ gauge group in the remaining part of the article. }:
\be\label{poten}
a^a_{\mu}= (a^3_{\mu}, a_{\mu}, a^{-}_{\mu}), ~~~~a_{\mu}=  a^{1}_{\mu} +i  a^{2}_{\mu} , 
~~~~a^-_{\mu}=   a^{1}_{\mu} -i  a^{2}_{\mu}.
\ee
For the neutral and charged components we will get: 
\beqa
 -\partial_{\mu} ( \partial_{\mu} a^{3}_{\nu} - \partial_{\nu} a^{3}_{\mu})  = \lambda a^{3}_{\nu},~~~~
 -\nabla_{\mu}( \nabla_{\mu} a_{\nu} - \nabla_{\nu} a_{\mu}) + i g F_{\mu\nu} a_{\mu} = \lambda a_{\nu},
\eeqa
where $\nabla_{\mu} = \partial_{\mu}  -  {1\over 2} i g F_{\mu\nu} x_{\nu}$. Taking into account that
$
[\nabla_{\mu},\nabla_{\nu}] = ig F_{\mu\nu} \nn
$
and imposing the background gauge fixing condition on the quantum field 
$
\nabla^{ab}_{\mu} a^{b}_{\mu} = 0
$
\cite{Batalin:1976uv}
for the charged components we will have: 
\be\label{eigenva}
 -\nabla_{\mu} \nabla_{\mu} a_{\nu}  + 2 i g F_{\mu\nu} a_{\mu} = \lambda a_{\nu}~.
\ee
For the vacuum field (\ref{interpol}), (\ref{backfields}) we will get \cite{Leutwyler:1980ev,  Leutwyler:1980ma, Savvidy:2022jcr}
\beqa\label{theh0}
H_0=-\nabla_{\mu} \nabla_{\mu} 
=  -\partial^2_1- \partial^2_2 + ig H (  x_2 \partial_1 -  x_1 \partial_2) + {g^2 \over 4} H^2 ( x^2_2+  x^2_1)\nn\\
  - \partial^2_3- \partial^2_0 + ig E (  x_0 \partial_3 -  x_3 \partial_0) + {g^2 \over 4} E^2 ( x^2_3+  x^2_0) 
\eeqa
and $H_0$ is a sum of isomorphic  oscillators in the $(1,2)$ and $(3,0)$ planes. The operators \cite{Leutwyler:1980ev, Leutwyler:1980ma}
\beqa\label{newoperators}
c_i = \partial_i + {g\over 2} H x_i,~~~~~c^+_i =  -\partial_i + {g\over 2} H x_i,~~~~~i=1,2\nn\\
d_j = \partial_j + {g\over 2} E x_j,~~~~~d^+_j =  -\partial_j + {g\over 2} E x_j,~~~~~j=3,0
\eeqa
allow to represent $H_0$  in the form
$
H_0=(c^+_1 +i c^+_2)(c_1 - i c_2)+(d^+_3 +i d^+_0)(d_3 - i d_0)+ g H + g E .
$
The eigenstates of the operator $H_0$ therefore are: 
\be\label{excitedsta}
\psi_{n m }= (c^+_1 +i c^+_2)^{n} (d^+_3 +i d^+_0)^{m} \psi_{00} = (g H)^n(x_1+i x_2)^n (g E)^m(x_3+ix_0)^m  \psi_{00},
\ee
where
$
 \psi_{00}(x)=e^{- {g H\over 4}(x^2_1 + x^2_2 )   } e^{ - {g E\over 4}( x^2_3  + x^2_0 )  }
$
and the corresponding eigenvalues $H_0 \psi_{n m } = \lambda_0 \psi_{n m}$ are: 
\be\label{H0spect}
\lambda_0 = (2n +1) gH + (2m +1) gE  .
\ee
All eigenstates have infinite degeneracy because the states 
\beqa\label{zeromodeswave}
~~~\psi_{ nm}(n_0, m_0)= (c^+_1-i c^+_2)^{n_0}(d^+_3 -i d^+_0)^{m_0} \psi_{nm}
\eeqa
have identical eigenvalues (\ref{H0spect}) and are indexed by two integers $n_0, m_0 =0,1,2...$. The eigenstates  of the charged field (\ref{poten}) are defined by the operator  
\be\label{Hoperator}
H_{\mu\nu}=- g_{\mu\nu }\nabla_{\lambda} \nabla_{\lambda}    - 2 i g F_{\mu\nu}
\ee
and have the following form:
$
b = a_1 +i a_2, ~~~b^- = a_1 - i a_2, ~~~ h= a_3 + i a_0,~~~ h^-= a_3 - i a_0.
$
Thus the eigenvalues  of the operator $H_{\mu\nu}$ will take the following form:
\beqa\label{spectr}
b^-:~~~~~\lambda_1=  (2n +1) gH + (2m +1) gE + 2 g H\nn\\
b~:~~~~~~\lambda_2 = (2n +1) gH + (2m +1) gE - 2 g H\nn\\
h^-:~~~~~\lambda_3 = (2n +1) gH + (2m +1) gE + 2 g E\nn\\
h~:~~~~~~\lambda_4 = (2n +1) gH + (2m +1) gE - 2 g E, 
\eeqa
where $n,m=0,1,2...$. The conjugate field  $a^-_{\mu}$ has the identical eigenvalues 
$\lambda_i$, $i=5,...,8$.  The negative eigenvalues appear in  $\lambda_2, \lambda_4$ when  $E \neq  H$ 
\cite{Nielsen:1978rm, Skalozub:1978fy,  Skalozub:1978fy, Ambjorn:1978ff, Nielsen:1978zg, Nielsen:1978nk, Ambjorn:1980ms} and only in the case of (anti)self-dual vacuum field \cite{Leutwyler:1980ev, Leutwyler:1980ma}
\be\label{selfdulitycon}
E=   H
\ee 
there are no negative/unstable eigenvalues in the spectrum (\ref{spectr}). \\
{\it {\bf 3}   Leutwyler zero mode chromons.}  It follows that when $n=m=0$, there is an infinite number of zero eigenvalues $\lambda_2 =  \lambda_4 = 0$  that appear due to the high symmetry of the self-dual field $E = H$ and the degeneracy (\ref{zeromodeswave}).  These are  {\it Leutwyler zero mode chromons} and they are linear combinations of (\ref{zeromodeswave}) ($b^{-}= a_1 - i a_2=0, ~h^{-}=a_3 - i a_0=0$):
\beqa\label{zero}
a_1(x) =i a_2(x)=  \sum_{n_0,m_0} \xi_{n_0 m_0} ~(c^+_1-i c^+_2)^{n_0}(d^+_3 -i d^+_0)^{m_0} \psi_{00}(x)\nn\\
a_3(x) =i a_0(x)=  \sum_{n_0,m_0} \eta_{n_0 m_0} ~(c^+_1-i c^+_2)^{n_0}(d^+_3 -i d^+_0)^{m_0} \psi_{00}(x),
\eeqa
where we introduced the zero-mode complex amplitudes (collective coordinates)  $\xi_{n_0 m_0}$ and $\eta_{n_0 m_0}$.  

On the zero mode eigenfunctions the quadratic form  (\ref{linear})  vanishes, $K_E=0$, and the amplitudes  can grow unbounded because the zero-mode amplitudes  are not suppressed by the exponent  $e^{- K_E }$.   It merely means that the integration in these directions is not Gaussian \cite{tHooft:1976snw}  and  the zero modes contribution to the effective Lagrangian  should be calculated  by integration over collective coordinates  (\ref{zero}) of nonlinearly interacting zero modes:
\be\label{zeromodemeasure}
\prod^{zero~modes}_{a, \mu} \CD a^{a}_{\mu}  =J \prod^{\CD eg}_{n_0,m_0=0} d\xi^{'}_{n_0 m_0} \wedge d\xi^{''}_{n_0 m_0}  \wedge d\eta^{'}_{n_0 m_0}  \wedge d\eta^{''}_{n_0 m_0}, 
\ee 
where $\xi_{n_0 m_0}=\xi^{'}_{n_0 m_0} +i \xi^{''}_{n_0 m_0}  $,  $\eta_{n_0 m_0}=\eta^{'}_{n_0 m_0} +i \eta^{''}_{n_0 m_0}  $ and $J$ is the  Jacobian factor\cite{tHooft:1976snw,Fujikawa:1979ay}.  The cubic self-interaction term (\ref{nonlinear}) vanishes in the zero mode directions (\ref{zero}):
\be\label{cubic1}
V^{(3)}_E=- g  \int  d^4 x \ \epsilon^{abc} a^{b}_{\nu}a^{c}_{\mu} \nabla^{ad}_{\mu} a^{d}_{\nu}=0.
\ee
The quartic interaction term is not vanishing  and the path integral has the following form:
\beqa\label{zeromodepartionfunc}
Z^{zero~modes}&=&N \int \exp{\Big[- {g^2 \over 4} \int d^4x  (\epsilon^{abc}a^{b}_{\mu}a^{c}_{\nu})^2    ~  \Big]} ~\prod^{zero~modes}_{a, \mu} \CD a^{a}_{\mu},
\eeqa
where the integration is over all interacting zero modes.  The  zero mode partition function (\ref{zeromodepartionfunc}), if expanded in the coupling constant $g^2$, will generate an infinite number of multi-loop diagrams, and it seems impossible to calculate them. Nevertheless the exact calculation of the zero mode partition function is possible because the field strength dependence within the path integral (\ref{zeromodepartionfunc}) can be factorised. 

Let us first consider the lowest state (\ref{spectr})  when $n=m=n_0=m_0=0$ in (\ref{excitedsta}) and (\ref{zeromodeswave}).   The solution (\ref{zero}) corresponding to the lowest state with $n_0=m_0=0$ is:
\beqa\label{zeroditections}
&a^1_1 = \xi^{'} \psi_{00},~~&a^2_1 = \xi^{''} \psi_{00},~~~~a^3_1=0\nn\\
&a^1_2 = \xi^{''} \psi_{00},~~&a^2_2 = -\xi^{'} \psi_{00},~~a^3_2=0\nn\\
&a^1_3 = \eta^{'} \psi_{00},~~&a^2_3 = \eta^{''} \psi_{00},~~~~a^3_3=0\nn\\
&a^1_0 = \eta^{''} \psi_{00},~~&a^2_0 = -\eta^{'} \psi_{00},~~a^3_0=0,
\eeqa
where $\xi_{00} = \xi^{'}  +i  \xi^{''}  $ and $\eta_{00} = \eta^{'}  +i  \eta^{''}  $.  On the zero mode chromon   (\ref{zeroditections}) the  quartic term will take the following form:
\be\label{quartic}
V^{(4)}_E= {g^2 \over 4} \int d^4x (\epsilon^{abc}a^{b}_{\mu}a^{c}_{\nu})^2
=  {g^2 \over 2}(\xi^{' 2}+\xi^{'' 2}+\eta^{' 2}+\eta^{'' 2})^2 \int d^4x  \vert \psi_{00}(x) \vert^4
\ee
and after integration over the space coordinates in (\ref{quartic}) the  partition function will take the following form:
\beqa
Z^{zero~mode}_{00}&=&  \mu^4 \int \exp{\Big[- {g^2 \over 2}  \Big({g H  \over 4 \pi}\Big)^2   (\xi^{' 2}+\xi^{'' 2}+\eta^{' 2}+\eta^{'' 2})^2   \Big] }~ d \xi^{' } d \xi^{'' } d \eta^{' } d \eta^{'' }.
\eeqa 
Introducing the dimensionless collective coordinates  $(\xi,\eta) \rightarrow (\xi,\eta)/(gH)^{1/2}$ one can factorise the field strength dependence in the path integral and we will get
\beqa
Z^{zero~mode}_{00}&=&   \Big({ \mu^2 \over g H } \Big)^2     \int \exp{\Big[- {g^2 \over 32}   (\xi^{' 2}+\xi^{'' 2}+\eta^{' 2}+\eta^{'' 2})^2   \Big] }~ d \xi^{' } d \xi^{'' } d \eta^{' } d \eta^{'' }= N_{00} \Big({ \mu^2 \over g H } \Big)^2.~~~~
\eeqa 
For the higher zero mode chromons  $\psi_{00}(n_0, m_0)  $ (\ref{zeromodeswave}) we will have 
\beqa\label{individualzero}
Z^{zero~mode}_{n_0 m_0}&=& \Big({ \mu^2 \over g H } \Big)^2     \int \exp{\Big[- {g^2 \over 32}  {\Gamma(n_0 +1/2) \Gamma(m_0 +1/2) \over \pi \Gamma(n_0+1) \Gamma(m_0+1)}  (\vert \xi_{n_0 m_0} \vert^2 +\vert \eta_{n_0 m_0} \vert^2)^2   \Big] }~ d^2 \xi_{n_0m_0} d^2 \eta_{n_0m_0}\nn\\
&=&  \Big({ \mu^2 \over g H } \Big)^2 \times N_{n_0 m_0}.     
\eeqa
In order to calculate the contributions of  self-interacting zero-modes  one should evaluate  the product $J \prod^{\CD eg}_{n_0=0,m_0=0} Z^{zero~mode}_{n_0 m_0}$ by taking into account the degeneracy (\ref{chromondege}) of the zero modes (\ref{zeromodeswave}):
\beqa\label{zeromodepro1}
Z_{zero~modes}&=& J \prod^{\CD eg}_{n_0=0,m_0=0} N_{n_0 m_0}  \Big( {  \mu^4 \over g^2 H^2 } \Big)   =
J  N  \Big(  { \mu^4 \over g^2 H^2 }  \Big)^{ \CD eg} =J N  e^{  -   {g^2 H^2 \over 4 \pi^2 }   \ln   { g^2 H^2  \over  \mu^4 }   V_4      } ,
\eeqa
where $N=  \prod^{\CD eg}_{n_0=0,m_0=0} N_{n_0 m_0}  $.  In the limit  $T \rightarrow \infty$  we have  $Z^{}_{zero~modes}  \rightarrow  e^{-\CL_{eff} V_4} $  and for the zero mode contribution (\ref{zeromodepro1}) to the effective Lagrangian  we will get
\be\label{zeromodenonlinear1}
\CL_{ zero~modes} =  {g^2 H^2 \over 4 \pi^2 }    \ln   {  g^2 H^2  \over  \mu^4 }   .
\ee
The contribution of the positive/stable modes has the following form \cite{Savvidy:2022jcr}:
 \beqa\label{sgablezeromode}
  \CL^{(1)}_{positive~modes} 
&= &  -{ g^2 H^2 \over 48 \pi^2}  \Big[  \ln {2 g^2 H^2 \over \mu^4} -1\Big ]. 
\eeqa
The effective Lagrangian is a sum of positive/stable modes (\ref{sgablezeromode}) and  zero modes chromons (\ref{zeromodenonlinear1}):  
\be\label{totaleffLag}
\CL^{eff}_E  =  H^2 + \CL^{(1)}_{positive~modes} + \CL_{ zero~modes} =H^2 + {11 g^2 H^2 \over 48 \pi^2}  \Big[  \ln {2 g^2 H^2 \over \mu^4} -1\Big]. 
\ee
There is a perfect consistency between this result  and the one obtained in a physical  space-time \cite{Savvidy:1977as,Savvidy:2019grj}:
\be\label{minlagra}
\CL_{YM}  
=- \CF  - {11 g^2 \CF  \over 48 \pi^2}  \Big[  \ln {2 g^2 \CF \over \mu^4} -1\Big], ~~~~~~~~~~~\CF= {1\over 4} G^a_{\mu\nu}G^a_{\mu\nu} >  0. 
\ee
As it was already mentioned, the electric field in a physical space-time  $\vec{E}$  corresponds to an imaginary  electric field     in Euclidian space $ \vec{E} \rightarrow  i \vec{E}$,     therefore one can map the gauge invariant operators as
\be
 \CF   = {1\over 2} (\vec{H}^2 -\vec{E}^2)  ~~\Rightarrow ~~   {1\over 2} (\vec{H}^2 +\vec{E}^2)_E = \CF_E
\ee
and the effective Lagrangian (\ref{minlagra})  transforms  into   the Euclidean  effective Lagrangian (\ref{totaleffLag}).   \\
{\it {\bf 4} Prefactor $N$. }  Our main concern now is the investigation of a possible field strength dependence  of the prefactor  $N$ and the Jacobian factor $J$ in the formula (\ref{zeromodepro1}) that can alter the effective Lagrangian (\ref{totaleffLag}). The integration over the zero mode collective coordinates is reminiscent  to the integration technique developed   by Feynman \cite{Feynman:1965}, t'Hooft \cite{tHooft:1976snw} and  Fujikawa \cite{Fujikawa:1979ay} and will be used to investigate the field strength dependence of the prefactor $N$ and Jacobian factor $J$. The prefactor $N$ is defined as in  (\ref{zeromodepro1})
\be\label{product}
N= \prod^{\CD eg}_{n_0=0,m_0=0} N_{n_0 m_0} ,
\ee 
 where from  (\ref{individualzero}) we have 
\beqa
 N_{n_0 m_0} =   \int \exp{\{- {g^2 \over 32}  {\Gamma(n_0 +1/2) \Gamma(m_0 +1/2) \over \pi \Gamma(n_0+1) \Gamma(m_0+1)}  (\vert \xi_{n_0 m_0} \vert^2 +\vert \eta_{n_0 m_0} \vert^2)^2   \} }~ d^2 \xi_{n_0m_0} d^2 \eta_{n_0m_0}. 
\eeqa
The prefactor $N$ can have a field strength dependence because  the degeneracy $\CD eg$ (\ref{chromondege}) of the eigenstates  (\ref{zeromodeswave}) had  a field strength dependent and is of the following form \cite{Savvidy:2022jcr}:
\be\label{chromondege} 
\CD eg =N_0 M_0 = {g H \over 2 \pi } A_{12}  {g E \over 2 \pi } A_{03} =  {g^2  E  H\over 4 \pi^2 } V_4 ,
\ee 
where 
\be
N_0  = {g H \over 2 \pi } A_{12}, ~~~~M_0  = {g E \over 2 \pi } A_{30}.
\ee
The $A_{12}$ is two-dimensional area on the plane $(x_1,x_2)$, while the
$A_{30}$ is two-dimensional area on the plane $(x_3,x_0)$ and the 4-volume is $V_4=A_{12} A_{30}$.  Rescaling the collective coordinates $(\xi, \eta)$ in the $N_{n_0 m_0}$ one can get alternative expression for $N_{n_0 m_0}$  with a factorised field strength dependence 
\beqa
 N_{n_0 m_0} &=& {32 \pi  \over g^2  }  {\Gamma(n_0+1) \Gamma(m_0+1) \over \Gamma(n_0 +1/2) \Gamma(m_0 +1/2) } \times \int e^{- (\vert \xi_{n_0 m_0} \vert^2 +\vert \eta_{n_0 m_0} \vert^2)^2   } d^2 \xi_{n_0m_0} d^2 \eta_{n_0m_0} \nn\\
 \nn\\
 &=&{32  \pi  \over g^2  }  {\Gamma(n_0+1) \Gamma(m_0+1) \over \Gamma(n_0 +1/2) \Gamma(m_0 +1/2) }  \times I , 
\eeqa
where 
\be
I= \int \exp{\{ - (\vert \xi_{n_0 m_0} \vert^2 +\vert \eta_{n_0 m_0} \vert^2)^2   \} }~ d^2 \xi_{n_0m_0} d^2 \eta_{n_0m_0} .\nn
\ee 
Evaluating  the product in (\ref{product})  for the   prefactor $N$ we will get: 
\beqa\label{prefactor1}
N&=& \prod^{\CD eg}_{n_0,m_0} N_{n_0 m_0} = \prod^{\CD eg}_{n_0,m_0}{32  \pi  I \over g^2  }  {\Gamma(n_0+1) \Gamma(m_0+1) \over \Gamma(n_0 +1/2) \Gamma(m_0 +1/2) }  =\nn\\
&=&\Big(  {2 e^3 \pi^6  \over A^{36} }  \Big)^{{N_0 +M_0 \over 24}}  \Big({32  \pi  I \over g^2  } \Big)^{N_0 M_0}
\Big[{G(N_0 +2)\over G(N_0 +3/2)}\Big]^{M_0}
\Big[{G(M_0 +2)\over G(M_0 +3/2)}\Big]^{N_0} 
\eeqa
where $A$ is the Glaisher constant and $G(n)$ is the Barnes function.  Because the contribution to the effective Lagrangian should scale with 4-volume $N_0M_0  \rightarrow V_4  $  we have to extract from the asymptotic behaviour of the prefactor $N$ only the terms that are proportional to the product $N_0M_0$ (\ref{chromondege}). Only the last two terms scale with the 4-volume and can contribution to the effective Lagrangian.  Using asymptotic  behaviour of the Garnes function at $N_0 \rightarrow \infty$, for the given ratio  one can get:
\be
{G(N_0 +2)\over G(N_0 +3/2)} \rightarrow  (2 \pi)^{1/4} ~e^{-{N_0 \over 2}    + ({3\over 8} + {N_0\over 2}) \log{N_0}}
\ee
and for the product in (\ref{prefactor1})
\beqa\label{prefactor2}
  \Big[{G(N_0 +2)\over G(N_0 +3/2)}\Big]^{M_0}
\Big[{G(M_0 +2)\over G(M_0 +3/2)}\Big]^{N_0}  &\rightarrow&    ( 2 \pi )^{{N_0+M_0\over 4}}~e^{ {3\over 8} ( M_0 \log{N_0}   + N_0 \log{M_0} )}  ~e^{-N_0 M_0    + {N_0 M_0 \over 2}  \log{(N_0 M_0)} } .~~~~~~
\eeqa
The last exponent has the terms that scales with  a  4-volume, thus 
\beqa
&&N \propto    ~e^{-N_0 M_0    + {N_0 M_0 \over 2}  \log{N_0 M_0} }  =
 ~e^{ 
    +{g^2  H^2 \over 8 \pi^2 } \Big( \log{ g^2  H^2 \over \mu^4 } -2 \Big) V_4  } 
\eeqa
and the contributions of the   prefactor $N$ to the effective Lagrangian is 
\be\label{Ncontribution}
\CL^{prefactor}_{eff}= - {g^2  H^2 \over 8 \pi^2 }   \log{ 2 g^2  H^2 \over \mu^4 } .
\ee
We  turn to the investigation of the integration measure and of the corresponding Jacobian factor. \\
{\it {\bf 5}    Jacobian J.} 
The   Leutwyller zero mode chromons  (\ref{zero})  can be represented in the following form:
\beqa\label{zero3}
a_1(x) =i a_2(x)=  \sum_{n_0,m_0} \xi_{n_0 m_0} ~(x_1-i x_2)^{n_0}(x_3 -i x_0)^{m_0} \psi_{00}(x)\nn\\
a_3(x) =i a_0(x)=  \sum_{n_0,m_0} \eta_{n_0 m_0} ~(x_1-i x_2)^{n_0}(x_3 -i x_0)^{m_0} \psi_{00}(x),
\eeqa
where for the self-dual field (\ref{selfdulitycon})
$$
 \psi_{00}(x)=e^{- {g H\over 4}(x^2_1 + x^2_2 + x^2_3  + x^2_0 )  } \equiv e^{- {g H\over 4} x^2   }.
$$
Introducing dimensionless coordinates $y_{\mu}= x_{\mu}/(g H)^{1/2}$ one can represent the gauge field in the form
\beqa\label{zero2}
a_1(y) =i a_2(y)=  \sum_{n_0,m_0} (g H)^{-{n_0 +m_0\over 2}}  \xi_{n_0 m_0}  ~(y_1-i y_2)^{n_0}(y_3 -i y_0)^{m_0} \psi_{00}(y)\nn\\
a_3(y) =i a_0(y)=  \sum_{n_0,m_0} (g H)^{-{n_0 +m_0\over 2}}  \eta_{n_0 m_0}   ~(y_1-i y_2)^{n_0}(y_3 -i y_0)^{m_0} \psi_{00}(y).
\eeqa
Let us divide the large volume $V_4$ in the Euclidean space into small 4-cubes $v_i$ (i=1,..,$N_0 M_0$) and approximate the gauge fields inside a small volume $v_i$ to be a constant, then the gauge fields in $V_4$  will have the following form:
\beqa\label{zero1}
&&a_1(i) =i a_2(i)=  \sum_{n_0,m_0}  (g H)^{-{n_0 +m_0\over 2}}  ~K_{n_0m_0}(i) ~ \xi_{n_0 m_0} \nn\\
&&a_3(i) =i a_0(i)=  \sum_{n_0,m_0} (g H)^{-{n_0 +m_0\over 2}} ~K_{n_0m_0}(i)  ~\eta_{n_0 m_0} , ~~~~~~i=1,...,N_0M_0,
\eeqa
where the kernel is
\be
K_{n_0m_0}(i) = ~(y_1(i)-i y_2(i))^{n_0}  ~(y_3(i) -i y_0(i))^{m_0} ~ \psi_{00}(y_i).
\ee
The part of the integral measure that include the field $a_1(i)$  will have the following form: 
\beqa
&\prod^{N_0M_0}_{i=1} d a_1(i) \wedge d a^*_1(i) = \\
&\sum_{n_0,m_0} (g H)^{-{n_0 +m_0\over 2}} K_{n_0m_0}(1)   d \xi_{n_0 m_0}\wedge d \xi^*_{n_0 m_0}   
\wedge ...\wedge 
  \sum_{n_0,m_0}  (g H)^{-{n_0 +m_0\over 2}}K_{n_0m_0}(N_0M_0)   d \xi_{n_0 m_0} \wedge d \xi^*_{n_0 m_0} .  \nn
  \eeqa
Opening the wedge product we will get the  determinant of a large matrix $\Delta$ of the size $N_0M_0 \times  N_0M_0$.  On each rows of the  $\Delta$ matrix there are identical field dependent factors  $(g H)^{-{n_0 +m_0\over 2}}$ and the matrix can be written in the following "symbolic"  form:
\be
\label{eqmatrix}
\Delta =  
   \begin{pmatrix}  
      (g H)^{-{0+0\over 2}} & ...  & ...  & ...  & ... &... & ...  \\
      ... & (g H)^{-{1 +0\over 2}} & ... & ... & ... &...& ... \\
      ... & ... & (g H)^{-{0 +1\over 2}} & ... & ... &...& ... \\
      ... & ... & ... & (g H)^{-{1 +1\over 2}} &   ... &...& ... \\
      ...&...&...&...&...&...&...\\
      ... & ... & ... &  ... & ... & ... & (g H)^{-{N_0 +M_0\over 2}}
   \end{pmatrix}.
\ee 
The field strength dependence can be extracted from each row and the Jacobian factor become proportional to the following product 
\be
J_1= \Det ~\Delta = \Lambda_1 ~\prod^{N_0,M_0}_{n_0,m_0 }  (g H)^{-{n_0 +m_0\over 2}}= \Lambda_1 ~
(g H)^{-{1\over 4} N_0 M_0 (N_0+ M_0+2)     },
\ee
where $\Lambda_1$ is the determinant of the residual matrix that is now without any field strength dependent factors. The full zero mode integration measure now will take the following form:
\be
\prod^{N_0M_0}_{i=1} d a_1(i) \wedge d a^*_1(i) \wedge  d a_3(i) \wedge  d a^*_3(i) 
 = J  \prod^{N_0,M_0}_{n_0,m_0 }  d \xi_{n_0 m_0}\wedge d \xi^*_{n_0 m_0}   
\wedge ...\wedge  d \eta_{n_0 m_0} \wedge d \eta^*_{n_0 m_0},
\ee
where 
\be
J = J_1 J_3   = \Lambda_1 \Lambda_3~ (g H)^{-{1\over 2} N_0 M_0 (N_0+ M_0+2) }  
\ee
Only  the $N_0 M_0$ term scales as 4-volume $V_4 $,  thus the part of Jacobian $J$ that contribute to the effective Lagrangian is proportional to the following expression 
\be
J ~~ \propto ~~   (g H)^{ - N_0 M_0    }= e^{-{g^2 H^2 \over 8 \pi^2} \log{ 2 g^2  H^2 \over \mu^4 } V_4} 
\ee
and the corresponding part of the  effective Lagrangian is 
\be\label{Jcontribution}
\CL^{Jacobian}_{eff}= + {g^2  H^2 \over 8 \pi^2 }   \log{ 2 g^2  H^2 \over \mu^4 } .
\ee
Summing the contributions to the effective Lagrangian  that appear  from the partition function prefactor $N$ (\ref{Ncontribution}) and Jacobian factor $J$ (\ref{Jcontribution}) one can see that these contributions  cancel each other 
$$
\CL^{prefactor}_{eff} + \CL^{Jacobian}_{eff} =0
$$ 
and therefore do not alter the expression for the  (anti)self-dual effective Lagrangian (\ref{totaleffLag}).  \\
{\it {\bf 6}   Degeneracy of  the Yang-Mills vacuum state.} 
In a more general case of covariantly-constant  chromomagnetic vacuum fields  \cite{Duff:1975ue,  Batalin:1976uv, PhDTheses},  instead of the zero mode chromons appear a plethora of negative modes displaying an apparent instability  \cite{Nielsen:1978rm, Skalozub:1978fy, Ambjorn:1978ff, Nielsen:1978zg, Nielsen:1978nk, Ambjorn:1980ms}. Generalisation of  the calculations that were advocated earlier  by Ambjorn, Nielsen, Olesen \cite{ Ambjorn:1978ff, Nielsen:1978zg, Nielsen:1978nk, Ambjorn:1980ms},  Flory \cite{Flory:1983td}  and other authors \cite{ Leutwyler:1980ev, Leutwyler:1980ma,  Minkowski:1981ma,    Savvidy:2019grj, parthasarathy,Kay1983, kumar,  Cho:1979nv,Pak:2017skw, Kondo:2013aga,Savvidy:2021ahq}  allows to obtain  the effective Lagrangian that includes contributions of positive/stable and negative/unstable modes and to demonstrate that the effective Lagrangian  is without imaginary terms \cite{Savvidy:2022jcr}.  All these vacuum fields are stable and indicate that the Yang-Mills vacuum is a highly degenerate  quantum state.

A larger class of  alternative vacuum fields have been considered in the literature as well \cite{Baseyan,  Natalia,   SavvidyPlanes,  Cho:1979nv,  Pak:2017skw, Kondo:2013aga}, and some of these vacuum fields  expose a natural chaotic behaviour \cite{Savvidy:2020mco}.   In this respect one should ask whether there exist physical systems that have high degeneracy of the vacuum state. Turning to the statistical spin systems, one can observe that the classical 3D Ising system has a double degeneracy of the vacuum state and of all its excited states. It is this symmetry that allows to construct a dual gauge invariant representation  of the 3D Ising model \cite{Wegner:1971app}. The extensions of the 3D Ising model that have a direct ferromagnetic  and  one quarter  of the next  to nearest  neighbour  antiferromagnetic interaction constructed in \cite{Savvidy:1993ej}, as well as a model with a zero intersection coupling constant ($k=0)$ \cite{Savvidy:1993sr, Savvidy:1994sc,Pietig:1996xj, Pietig:1997va}, have {\it exponential degeneracy of the vacuum state}.  In recent publications this symmetry was referred as the {\it subsystem symmetry} \cite{Vijay:2016phm}. This higher symmetry allows to construct the dual representations of the same systems  and in various  dimensions \cite{Savvidy:1993sr, Savvidy:1994sc,Savvidy:1994tf}.  As a  consequence of the high degeneracy of the vacuum state,  these systems have rich physical properties, including a glass behaviour \cite{Lipowski, Sherrington}, exotic fracton excitations \cite{Vijay:2016phm}, and it was suggested that they can be used as high density memory devices \cite{Savvidy:2000zq}.


\begin{thebibliography}{100}

\bibitem{Leutwyler:1980ev}
  H.~Leutwyler,
 \emph{Vacuum Fluctuations Surrounding Soft Gluon Fields,}
  Phys.\ Lett.\  {\bf 96B} (1980) 154.
  doi:10.1016/0370-2693(80)90234-8
   
   
\bibitem{Leutwyler:1980ma}
  H.~Leutwyler,
 \emph{Constant Gauge Fields and their Quantum Fluctuations,}
  Nucl.\ Phys.\ B {\bf 179} (1981) 129.
  doi:10.1016/0550-3213(81)90252-2


\bibitem{Minkowski:1981ma}
P.~Minkowski,
 \emph{On the Ground State Expectation Value of the Field Strength Bilinear in Gauge Theories and Constant Classical Fields,}
Nucl. Phys. B \textbf{177} (1981), 203-217
doi:10.1016/0550-3213(81)90388-6
   

\bibitem{Savvidy:2022jcr}
G.~Savvidy,
 \emph{Stability of Yang Mills Vacuum State,}
[arXiv:2203.14656 [hep-th]].

  
\bibitem{Savvidy:1977as}
  G.~K.~Savvidy,
 \emph{Infrared Instability of the Vacuum State of Gauge Theories and Asymptotic Freedom,}
  Phys.\ Lett.\  {\bf 71B} (1977) 133.
  doi:10.1016/0370-2693(77)90759-6


\bibitem{Matinyan:1976mp}
  S.~G.~Matinyan and G.~K.~Savvidy,
 \emph{Vacuum Polarization Induced by the Intense Gauge Field,}
  Nucl.\ Phys.\ B {\bf 134} (1978) 539.
  doi:10.1016/0550-3213(78)90463-7

\bibitem{Duff:1975ue}
  M.~J.~Duff and M.~Ramon-Medrano,
 \emph{On the Effective Lagrangian for the Yang-Mills Field,}
  Phys.\ Rev.\ D {\bf 12} (1975) 3357.
https://doi.org/10.1103/PhysRevD.12.3357


\bibitem{Batalin:1976uv}
  I.~A.~Batalin, S.~G.~Matinyan and G.~K.~Savvidy,
 \emph{Vacuum Polarization by a Source-Free Gauge Field,}
  Sov.\ J.\ Nucl.\ Phys.\  {\bf 26} (1977) 214
   [Yad.\ Fiz.\  {\bf 26} (1977) 407].

\bibitem{PhDTheses}
  G.~K.~Savvidy,
 \emph{Vacuum Polarisation by Intensive Gauge Fields,} PhD 1977,\\
     \url{http://www.inp.demokritos.gr/~savvidy/phd.pdf}


\bibitem{Savvidy:2019grj}
G.~Savvidy,
 \emph{From Heisenberg\textendash{}Euler Lagrangian to the discovery of Chromomagnetic Gluon Condensation,}
Eur. Phys. J. C \textbf{80} (2020) no.2, 165
doi:10.1140/epjc/s10052-020-7711-6
[arXiv:1910.00654 [hep-th]].




\bibitem{tHooft:1976snw}
G.~'t Hooft,
 \emph{Computation of the Quantum Effects Due to a Four-Dimensional Pseudoparticle,}
Phys. Rev. D \textbf{14} (1976), 3432-3450
https://doi.org/10.1103/PhysRevD.14.3432


\bibitem{Feynman:1965}
R.~P.~Feynman  and A. R.~ Hibbs, 
 \emph{Quantum Mechanics and Path Integrals},
 McGraw-Hill,  New York 1965.
  
 
\bibitem{Fujikawa:1979ay}
K.~Fujikawa,
 \emph{Path Integral Measure for Gauge Invariant Fermion Theories,}
Phys. Rev. Lett. \textbf{42} (1979), 1195-1198
doi:10.1103/PhysRevLett.42.1195

\bibitem{Heisenberg:1935qt}
  W.~Heisenberg and H.~Euler,
\emph{Consequences of Dirac's theory of positrons,}
  Z.\ Phys.\  {\bf 98} (1936) 714.
     
\bibitem{Schwinger:1951nm}
  J.~S.~Schwinger,
\emph{On gauge invariance and vacuum polarization,}
  Phys.\ Rev.\  {\bf 82} (1951) 664.
  doi:10.1103/PhysRev.82.664

\bibitem{Coleman:1973jx}
  S.~R.~Coleman and E.~J.~Weinberg,
 \emph{Radiative Corrections as the Origin of Spontaneous Symmetry Breaking,}
  Phys.\ Rev.\ D {\bf 7} (1973) 1888.
https://doi.org/10.1103/PhysRevD.7.1888

\bibitem{Coleman}
 S. Coleman, \emph{The uses of Instantons,} Erice Lectures, 1977

\bibitem{Flory:1983td}
  C.~A.~Flory,
 \emph{Covariant Constant Chromomagnetic Fields And Elimination Of The One Loop Instabilities,}
Preprint, SLAC-PUB-3244,  http://www-public.slac.stanford.edu/sciDoc/docMeta.aspx?slacPubNumber=SLAC-PUB-3244; 
 https://lib-extopc.kek.jp/preprints/PDF/1983/8312/8312331.pdf

\bibitem{parthasarathy}
  D.~Kay, R.~Parthasarathy and K.~S.~Viswanathan
 \emph{Constant self-dual Abelian gauge fields and fermions 
 in SU(2) gauge theory,}
  Phys.\ Phys.\ D   {\bf 28} (1983) 3116-3120.

\bibitem{Kay1983}
D.~ Kay.~  \emph{Unstable modes, zero modes, and phase transitions in QCD}, Ph.D Thesis,
Simon Fraser University, August 1985.

\bibitem{kumar}
  D.~Kay, A.~Kumar and R.~Parthasarathy
 \emph{Savvidy Vacuum in SU(2) Yang-Mills Theory,}
  Mod.\ Phys.\  Lett.\ A {\bf 20} (2005) 1655-1662.



\bibitem{Nielsen:1978rm}
  N.~K.~Nielsen and P.~Olesen,
 \emph{An Unstable Yang-Mills Field Mode,}
  Nucl.\ Phys.\ B {\bf 144} (1978) 376.
  doi:10.1016/0550-3213(78)90377-2


\bibitem{Skalozub:1978fy}
  V.~V.~Skalozub,
 \emph{On Restoration of Spontaneously Broken Symmetry in Magnetic Field,}
  Yad.\ Fiz.\  {\bf 28} (1978) 228.

\bibitem{Ambjorn:1978ff}
  J.~Ambjorn, N.~K.~Nielsen and P.~Olesen,
 \emph{A Hidden Higgs Lagrangian in {QCD},}
  Nucl.\ Phys.\ B {\bf 152} (1979) 75.
  doi:10.1016/0550-3213(79)90080-4


\bibitem{Nielsen:1978zg}
  H.~B.~Nielsen,
 \emph{Approximate {QCD} Lower Bound for the Bag Constant $B$,}
  Phys.\ Lett.\  {\bf 80B} (1978) 133.
  doi:10.1016/0370-2693(78)90326-X  


\bibitem{Nielsen:1978nk}
N.~K.~Nielsen and P.~Olesen,
 \emph{Electric Vortex Lines From the {Yang-Mills} Theory,}
Phys. Lett. B \textbf{79} (1978), 304
doi:10.1016/0370-2693(78)90249-6


\bibitem{Ambjorn:1980ms}
  J.~Ambjorn and P.~Olesen,
 \emph{A Color Magnetic Vortex Condensate in QCD,}
  Nucl.\ Phys.\ B {\bf 170} (1980) 265.
  doi:10.1016/0550-3213(80)90150-9


\bibitem{Cho:1979nv}
Y.~M.~Cho,
\emph{A Restricted Gauge Theory,}
Phys. Rev. D \textbf{21} (1980), 1080
doi:10.1103/PhysRevD.21.1080236-3

\bibitem{Pak:2017skw}
D.~G.~Pak, B.~H.~Lee, Y.~Kim, T.~Tsukioka and P.~M.~Zhang,
 \emph{On microscopic structure of the QCD vacuum,}
Phys. Lett. B \textbf{780} (2018), 479-484
doi:10.1016/j.physletb.2018.03.040
[arXiv:1703.09635 [hep-th]].   

\bibitem{Kondo:2013aga}
K.~I.~Kondo,
 \emph{Stability of magnetic condensation and mass generation for confinement in SU(2) Yang-Mills theory,}
PoS \textbf{QCD-TNT-III} (2013), 020
doi:10.22323/1.193.0020
[arXiv:1312.0053 [hep-th]].  


\bibitem{Savvidy:2021ahq}
G.~Savvidy,
\emph{Gauge field theory vacuum and cosmological inflation without scalar field,}
Annals Phys. \textbf{436} (2022), 168681
https://doi.org/10.1016/j.aop.2021.168681
[arXiv:2109.02162 [hep-th]].




\bibitem{Baseyan}
G.~Baseyan, S.~ Matinyan and G.~ Savvidy,   
 \emph{Nonlinear plane waves in the massless Yang-Mills theory,} 
 Pisma Zh.\ Eksp.\ Teor.\ Fiz.\  {\bf 29}  (1979) 641-644 

 
\bibitem{Natalia}
S.~ Matinyan, G.~ Savvidy   and N.~Ter-Arutyunyan-Savvidi,
 \emph{Classical  Yang-Mills mechanics. Nonlinear colour oscillations,} 
 Zh.\ Eksp.\ Teor.\ Fiz.\  {\bf 80}  (1980) 830-838 
 
\bibitem{SavvidyPlanes}
G.~ Savvidy,   
 \emph{Classical  and Quantum mechanics of non-Abelian gauge fields,} 
 Nucl.\ Phys.\  {\bf 246}  (1984) 302-334 

   
\bibitem{Savvidy:2020mco}
G.~Savvidy,
\emph{Maximally chaotic dynamical systems,}
Annals Phys. \textbf{421} (2020), 168274
https://doi.org/10.1016/j.aop.2020.168274

\bibitem{Wegner:1971app}
F.~J.~Wegner,
\emph{Duality in Generalized Ising Models and Phase Transitions Without Local Order Parameters,}
J. Math. Phys. \textbf{12} (1971), 2259-2272
https://doi.org/10.1063/1.1665530

\bibitem{Savvidy:1993ej}
G.~K.~Savvidy and F.~J.~Wegner,
\emph{Geometrical string and spin systems,}
Nucl. Phys. B \textbf{413} (1994), 605-613
https://doi.org/10.1016/0550-3213(94)90003-5
[arXiv:hep-th/9308094 [hep-th]].


\bibitem{Savvidy:1993sr}
G.~K.~Savvidy and K.~G.~Savvidy,
\emph{Selfavoiding gonihedric string and spin systems,}
Phys. Lett. B \textbf{324} (1994), 72-77
https://doi.org/10.1016/0370-2693(94)00114-6
[arXiv:hep-lat/9311026 [hep-lat]].




\bibitem{Savvidy:1994sc}
G.~K.~Savvidy and K.~G.~Savvidy,
\emph{Interaction hierarchy,}
Phys. Lett. B \textbf{337} (1994), 333-339
https://doi.org/10.1016/0370-2693(94)90984-9
[arXiv:hep-th/9409030 [hep-th]].


\bibitem{Savvidy:1994tf}
G.~K.~Savvidy, K.~G.~Savvidy and F.~J.~Wegner,
\emph{Geometrical string and dual spin systems,}
Nucl. Phys. B \textbf{443} (1995), 565-580
https://doi.org/10.1016/0550-3213(95)00151-H
[arXiv:hep-th/9503213 [hep-th]].



\bibitem{Pietig:1996xj}
R.~Pietig and F.~J.~Wegner,
\emph{Phase transition in lattice surface systems with gonihedric action,}
Nucl. Phys. B \textbf{466} (1996), 513-526
https://doi.org/10.1016/0550-3213(96)00072-7
[arXiv:hep-lat/9604013 [hep-lat]].


\bibitem{Pietig:1997va}
R.~Pietig and F.~J.~Wegner,
\emph{Low temperature expansion of the gonihedric Ising model,}
Nucl. Phys. B \textbf{525} (1998), 549-570
https://doi.org/10.1016/S0550-3213(98)00342-3
[arXiv:hep-lat/9712002 [hep-lat]].


\bibitem{Lipowski}
A.~ Lipowski, D.~ Johnston, and D. ~Espriu, 
\emph{Slow dynamics of Ising models with energy barriers,}
 Phys. Rev. E \textbf{62} (2000)  3404.
https://doi.org/10.1103/PhysRevE.62.3404



\bibitem{Sherrington}
C.~Castelnovo, C.~ Chamon, and D.~Sherrington,
\emph{Quantum mechanical and information theoretic view on classical glass transitions,}
Phys. Rev. B \textbf{81} (2010) 184303 
https://doi.org/10.1103/PhysRevB.81.184303


\bibitem{Vijay:2016phm}
S.~Vijay, J.~Haah and L.~Fu,
\emph{Fracton Topological Order, Generalized Lattice Gauge Theory and Duality,}
Phys. Rev. B \textbf{94} (2016) no.23, 235157
https://doi.org/10.1103/PhysRevB.94.235157
[arXiv:1603.04442 [cond-mat.str-el]].

 
\bibitem{Savvidy:2000zq}
G.~K.~Savvidy,
 \emph{The System with exponentially degenerate vacuum state,}-
[arXiv:cond-mat/0003220 [cond-mat]].   


















 
\end{thebibliography}
\end{document}